# XORSAT: An Efficient Algorithm for the DIMACS 32-bit Parity Problem[*]

Jing-Chao Chen

School of Informatics, Donghua University, 1882 Yan-An West Road,
Shanghai, 200051, P. R. China
chen-jc@dhu.edu.cn
December 27,2006

**Abstract.** The DIMACS 32-bit parity problem is a satisfiability (SAT) problem hard to solve. So far, EqSatz by Li is the only solver which can solve this problem. However, This solver is very slow. It is reported that it spent 11855 seconds to solve a par32-5 instance on a Maxintosh G3 300 MHz. The paper introduces a new solver, XORSAT, which splits the original problem into two parts: structured part and random part, and then solves separately them with WalkSAT and an XOR equation solver. Based our empirical observation, XORSAT is surprisingly fast, which is approximately 1000 times faster than EqSatz. For a par32-5 instance, XORSAT took 2.9 seconds, while EqSatz took 2844 seconds on Intel Pentium IV 2.66GHz CPU. We believe that this method significantly different from traditional methods is also useful beyond this domain.

## 1 Introduction

The DIMACS 32-bit parity problem is considered as one of hard satisfiability (SAT) problems. Conducting research on this problem is of interest for finding efficient SAT algorithms, and for understanding better the complexity of SAT. A SAT instance is a propositional formula in conjunctive normal form (CNF), which is a conjunction of clauses, each of which is a disjunction of Boolean literals, where each literal is either a variable or the negation of a variable. The SAT problem is to determine for a given CNF formula, if there exists an assignment of truth values (1 or 0) to its variables for which each clause in that formula is true. This was the earliest NP-complete problem discovered by Cook [3]. Because it is simple yet hard, it has received an extensive attention.

In general, SAT solvers are divided into two categories: *complete* (called also *systematic*) and *incomplete* (*non-systematic*). An algorithm is said to be *complete* if for soluble problems, it is guaranteed to find a solution. Otherwise, it is said to be incomplete. Up to now, many SAT solvers have been developed, for example, GRASP [11], POSIT [7], Rel_Sat [2], ZChaff [13], BerkMin [8], Siege [14], SATELITE [5, 6], EqSatz [9, 10] and WalkSAT [12]. Most of them are complete and are based on the Davis-Putnam (DP) [4] backtrack search. Whether complete or incomplete, except for EqSatz, no SAT solver so far is suited for the DIMACS 32-bit Parity Problem, which is listed as Challenge 2 in the "Ten challenges in propositional reasoning and search" suggested by the AT&T researchers [15]. EqSatz can solve the challenge DIMACS 32-bit parity problem in reasonable time, but is very slow. Based on Li's experiments [10], the running time taken to solve a par32-5 instance was 11855 seconds on a Maxintosh G3 300 MHz. To our best knowledge, so far, this is the only solver that successfully solves this problem.

---

[*] This work was partially supported by the National Natural Science Foundation of China grant 60473013

The DIMACS parity problem originates from the minimal disagreement parity problem [1]. Selman et al. asserted that any algorithm solving it will have to do something significantly different from current methods [15]. Indeed, one will see that our algorithm is significantly different from the methods formerly known. In devising a new solver, we are aware that the instances of the DIMACS parity problem [16] were generated by random noises. And WalkSAT [12] is good at SAT instances generated uniformly at random, but not suited for structured SAT instances. In addition to random noise, the DIMACS parity problem contains some structured information. Therefore, in solving this problem, we split it into two parts: random and structured. The structured part is expressed by a set of XOR equations, which can be solved directly in a way similar to solving a system of linear equations. The random part is expressed by a CNF formula, which can be solved efficiently by a simplified WalkSAT. The resulting algorithm is very fast and can find a solution for a par32-5 instance in less than 3 seconds on an ordinary personal computer.

## 2 Extracting XOR Equations from a CNF Formula

In DIMACS suite [16], like other SAT instances, each instance of the parity problem is formulated as a CNF formula. The first task of our algorithm is to extract XOR equations from a CNF formula. To do this, we first transform some CNF clauses into ternary XOR equations. The notion of a ternary XOR equation here corresponds to the notion of a ternary equivalency clause in [10]. Let $\oplus$ stand for a XOR (exclusive-or) operation, i.e. equivalently modulo 2 arithmetic. A ternary XOR equation:

$$A \oplus B \oplus C = 1$$

is equivalent to 4 CNF clauses:

$A \vee \neg B \vee \neg C$,
$\neg A \vee B \vee \neg C$,
$\neg A \vee \neg B \vee C$

and $A \vee B \vee C$,

where $A$, $B$ and $C$ are a variable, and $\neg$ denotes the negation of a variable. Thus, we replace the 4 CNF clauses by an XOR equation. This substitution operation yields a large amount of ternary XOR equations for each instance of the parity problem. Then we merge the ternary XOR equations obtained into representative multi-nary XOR equations. The basic merging rule is that whenever two XOR equations share one variable, we merge them into one XOR equation. Notice, for the ease of operations, all ternary equations for merging are normalized into a standard form $A \oplus B \oplus C = c$, where $c$ is a constant equal to 0 or 1. That is, each equation has no the negation of a variable. The following is an example for merging. $A \oplus B \oplus C = 1$ and $C \oplus D \oplus F = 1$ are given, Applying the XOR operation in both sides, we have

$$A \oplus B \oplus C \oplus C \oplus D \oplus F = 1 \oplus 1.$$

Simplification yields our merged form:

$$A \oplus B \oplus D \oplus F = 0.$$

The merging implies the fact that if variable $C$ does not occur in other clauses, the two equations before merging and the equation after merging are equivalent. Continuing to merge it with other ternary equation yields a quinary equation. In general, merging an $n$-nary equation with a ternary equation yields an $(n+1)$-nary equation. We merge repeatedly with a ternary equation to obtain the longest possible equation. When ternary equations are exhausted, the merging process terminates. Here, there exists such a problem: how to determine the first ternary XOR equation for each required equation. Our solution is to select a ternary XOR equation with two frequently used variables as the first ternary XOR equation. According this rule, we extracted a set of 32 XOR equations on 48 variables for each instance of the DIMACS 16-bit Parity Problem. As an example, the following is an equation extracted from a par16-2 instance.

$$X_4 \oplus X_{550} \oplus X_{552} \oplus X_{554} \oplus X_{556} \oplus X_{557} \oplus X_{558} \oplus X_{560} \oplus X_{17} = 0.$$

In this equation, except for $X_{17}$, all variables are the most frequently used ones. For each instance of the DIMACS 32-bit Parity Problem, we extracted a set of 64 XOR equations on 96 variables.

In the system of linear equations, there is an approach called Gauss-Jordan elimination, which find a solution by converting a matrix into a reduced echelon form using elementary row operations. The basic steps are as follows: first create leading 1s, and then eliminate terms that are not selected as leading 1s so that columns containing leading 1s have only 0s above and below the leading 1. This is done in column by column starting with the first column. The Gauss-Jordan elimination is also suited for the system of XOR equations. Using this approach, we can convert a set of XOR equations in the parity problem into the following form:

$$(XE) \begin{cases} x_1 = a_{10} \oplus a_{11} y_1 \oplus a_{12} y_2 \oplus \ldots \oplus a_{1n} y_n \\ x_2 = a_{20} \oplus a_{21} y_1 \oplus a_{22} y_2 \oplus \ldots \oplus a_{2n} y_n \\ \vdots \\ x_n = a_{n0} \oplus a_{n1} y_1 \oplus a_{n2} y_2 \oplus \ldots \oplus a_{nn} y_n \end{cases}$$

$$(ZE) \begin{cases} z_1 = a_{(n+1)0} \oplus a_{(n+1)1} y_1 \oplus a_{(n+1)2} y_2 \oplus \ldots \oplus a_{(n+1)n} y_n \\ z_2 = a_{(n+2)0} \oplus a_{(n+2)1} y_1 \oplus a_{(n+2)2} y_2 \oplus \ldots \oplus a_{(n+2)n} y_n \\ \vdots \\ z_m = a_{(n+m)0} \oplus a_{(n+m)1} y_1 \oplus a_{(n+m)2} y_2 \oplus \ldots \oplus a_{(n+m)n} y_n \end{cases}$$

where $a_{ij}$ ($i = 0,1,2,\ldots,n+m$, $j=0,1,2,\ldots,n$) is a constant equal to 0 or 1, $x_i$ ($i=0,1,2,\ldots,n$) is a frequently used variable, and $y_i$ ($i=0,1,2,\ldots,n$) and $z_i$ ($i=0,1,2,\ldots,m$) are a usual variable. The values of $n$ and $m$ depend on actual problems. For the DIMACS 32-bit parity problem, both *n* and *m* are 32. The solution of the system depends on $y_1$, $y_2,\ldots,y_n$, which will be solved by a variant of WalkSAT plus an enumeration.

## 3   A Variant of WalkSAT

```
procedure WalkSAT(F, maxTries, Cutoff, heuristic)
  for try := 1 to maxTries do
     V:=randomly chosen assignment of the variables in F
     for i := 1 to Cutoff do
        if V satisfies F then return a
        c:=randomly selected clause which is unsatisfied under a
        x:=variable in c selected according to heuristic
        V:=V with x flipped;
     end for;
  end for;
  return "no solution found"
end WalkSAT
```

**Fig. 1.** The WalkSAT algorithm.

WalkSAT [12] is a stochastic local search (SLS) algorithm, which is considered as one of the fast solvers. This algorithm can outperform the best systematic search algorithms on a number of domains, and solve efficiently large and hard SAT instances. Its drawback is *incomplete*. The algorithm views the set of all truth assignments for variables appearing in the CNF formula *F* as a state. Each local search step changes at most the value of a variable in the state. Such a change is called a *variable flip*. The basis of each change is the value of the objective function. The simplest objective function is defined as the number of clauses which are unsatisfied under a given state. To avoid getting trapped in local minima of the objective function, the popular idea is to perform random hill-climbing on the objective function. The WalkSAT algorithm for solving a CNF formula *F* is outlined in Fig. 1.

What we want to solve with WalkSAT is the random part, which excludes the structured component. To do this, it is not necessary to utilize fully the power of WalkSAT, since we note that the random part of the parity problem is actually easier. We therefore simplify WalkSAT. The first idea as simplification is to remove the random generator. This has two advantages. One is that it can speed up the SAT solver. The other is that it can easily recover the previous experiment. The second idea is to reduce the number of tries. We try only two initial assignments: one in which all variables are set to false and one in which all variables are set to true. The third idea is to reduce the maximum number of flips. We set the maximum number of flips to double the number of clauses. This is usually very small. For par32 instances, it is about 3500. Fig. 2 shows a simple WalkSAT algorithm based on these ideas. Compared to the original WalkSAT in Fig. 1, this procedure removes parameter *heuristic*. So far, many heuristics for WalkSAT have been developed, e.g. *best*, *TABU*, *Novelty*, *Novelty+*, *R-Novelty* and *R-Novelty+*. Based on our observation,

*Novelty+* is best suited for our purpose. Hence, we use it directly as a fixed parameter. The main idea of heuristic *Novelty+* is outlined as follows. Each variable is assigned to a score, which is defined as the difference between the number of makes and the number of breaks. The number of makes refers to the number of satisfied clauses the variable's flip would cause. And the number of breaks refers to the number of clauses which are currently satisfied but would become violated by the variable's flip. The variable with the maximal score is called the best variable. If the best variable in the selected clause has not been flipped the most recently, it is flipped. Otherwise, it is flipped every two flips, while in the other cases, the second-best variable is flipped. Here, we do not consider the case that the variable is flipped with a fixed probability $1-p$ (this is a point of the original *Novelty+*), since we do not want to use a random generator.

```
procedure sWalkSAT(F)
 for i:=1 to 2 do
   if i=1 then V:=assignment in which each x in F  is set to false
   if i=2 then V:=assignment in which each x in F  is set to true
   for j:=1 to 2 × #cluase do
      if V satisfies F then return V
      c:=in turn selected clause which is unsatisfied under a
      x:=variable in c selected according to heuristic novelty+
      V:=V with x flipped
    end for
   end for
   return "no solution found"
  end sWalkSAT
```

**Fig. 2.** A simple WalkSAT algorithm without any random generator.

## 4 The Algorithm for the Parity Problem

Fig. 3 shows a new SAT solver, *XORSAT*, in a pseudo-code. Because this new SAT solver contains the component of an XOR equation solver, it is named XORSAT. This solver consists mainly of a CNF simplification, XOR equation extraction, solving the random part and modifying the partial solution into a complete solution. The CNF simplification is used to reduce unit clauses and binary equivalency clauses. This step will delete the literals that are in the unit clauses and the equivalency clauses by a unit propagation procedure. For the details, see [10]. During the XOR equation extraction, we employ the notion of a *frequently used variable*, which is defined as a variable whose frequency of occurrence is greater than $\theta = 3\lceil \#clause / \#var \rceil + 2$, where $\#clause$ and $\#var$ denote the number of clauses and variables, respectively. Notice, for other problems, $\theta$ is not necessarily this value. Let $X$ be the set of frequently used variables. As described in Section 2, we can extract the set $E$ of XOR equations. Using the Gauss-Jordan elimination, we can convert the $E$ into

$XE = \{ x_i = a_{i0} \oplus a_{i1}y_1 \oplus a_{i2}y_2 \oplus \ldots \oplus a_{in}y_n \mid i=1,\ldots,n \}$

and $ZE = \{ z_{i-n} = a_{i0} \oplus a_{i1}y_1 \oplus a_{i2}y_2 \oplus \ldots \oplus a_{in}y_n \mid i=n+1,\ldots,n+m \}$.

```
procedure XORSAT(F)
  F:=simplified CNF formula of F
  Let X={x | the occurrence frequency of variable x ≥ θ}
  E:=a set of XOR equations extracted in F with X
  S:={s | clause s contains variables in X}
  V:=sWalkSat(F-S)
  Solving E with the Gauss-Jordan elimination yields
  XE={x_i = a_{i0}⊕a_{i1}y_1⊕ a_{i2}y_2 ⊕…⊕ a_{in}y_n | i=1,…,n}
  ZE={z_{i-n}= a_{i0}⊕a_{i1}y_1⊕ a_{i2}y_2 ⊕…⊕ a_{in}y_n | i=n+1,…,n+m}
  Let Y'={y_i' | y_i' is the truth assignment of y_i in V,1≤ i ≤ n}
  for each Y satisfying |Y-Y'| ≤ 3 do
      Calculating the XOR expressions in ZE with Y yields a solution $z_i'$ of $z_i$
      V:=UnitResolution(F-S,{y_1, y_2,…,y_n, z_1', z_2',…,z_m'})
      if V satisfies F-S then
          Calculating the XOR expressions in XE with Y yields a solution $x_i'$ of $x_i$
          return V+{x_1',x_2'…,x_n'}
      end if
  end for
  return "no solution found"
end XORSAT
```

**Fig. 3.** XORSAT: a SAT solver for the parity problem

Once $y_1, y_2, .., y_n$ are determined, $x_i$ and $z_i$ are easily determined. To compute $y_i$, we use *s*WalkSat to get a set $V$ of truth assignments satisfying the random part of the CNF formula $F$, which excludes clauses that contains variables in $X$. Of course, the $V$ contains the truth assignment of $y_i$. Let $Y'=\{y_1', y_2', .., y_n'\}$ be the truth assignment set in $V$. The truth assignments obtained by applying $Y'$ to $XE$ and $ZE$ is not necessarily the solution to the original problem. Based on our observation, the set of the truth assignments is very close to the final solution. In general, $Y'$ is at Hamming distance at most 3 away from $Y$ in the final solution. Therefore, for each truth assignment set $Y$ at Hamming distance at most 3 from $Y'$, we check whether it satisfies $F$. This can be done by computing $z_i$ with $Y$, and then getting the truth assignment of other variables with a unit resolution procedure. In Fig. 3, the notation $|Y–Y'| \leq 3$ denotes the Hamming distance between $Y$ and $Y'$ are at most 3. The unit resolution procedure in Fig. 3 works as the unit propagation in the usual DP procedure. Its task is to fix as many literals as possible and return them by making repeatedly such an obvious inference: assign a literal to one fixed value when all other literals in a clause are fixed but unsatisfy the clause. Assuming that each literal in $V$ is fixed, this procedure is described as follows.

```
procedure UnitResolution(F, V)
  while there exists a clause c in which only one literal x has not fixed under V, and
        to satisfy the c, the x needs be fixed to true.
      V := V ∪ {x}
  end while
  return V
```

In the above procedure, whether a literal $x$ is fixed can be done by testing whether $x \in V$, where $x$ is a variable or the negation of a variable.

## 5 Empirical evaluation

**Table 1.** Run time of *EqSATz* and *XORSAT* on the DIMACS 32-bit parity problem (in seconds)

| Instance | #var | #clause | *EqSATz* time | *XORSAT* time |
|---|---|---|---|---|
| par32-1-c | 1315 | 5254 | 251.3 | 0.110 |
| par32-2-c | 1303 | 5206 | 9.3 | 0.172 |
| par32-3-c | 1325 | 5294 | 1006.2 | 0.047 |
| par32-4-c | 1333 | 5226 | 152.0 | 1.531 |
| par32-5-c | 1339 | 5350 | 2203.8 | 0.640 |
| par32-1 | 3176 | 10277 | 181.5 | 0.094 |
| par32-2 | 3176 | 10253 | 44.0 | 0.422 |
| par32-3 | 3176 | 10297 | 2062.9 | 3.235 |
| par32-4 | 3176 | 10313 | 170.6 | 0.219 |
| par32-5 | 3176 | 10325 | 2844.6 | 2.922 |

We conducted our experiments under such a machine: Intel Pentium 4 CPU with speed of 2.66GHz. This machine is at least 4 times faster than Maxintosh G3 300 MHz used in Li's experiments [10], since for the same instance, say par32-5, using the same SAT solver, say *EqSATz*, the former took 2844.6 seconds, while the latter did 11855 seconds. Even so, *EqSATz* runs still very slow. We tested the run time of two SAT solvers: *EqSATz* and *XORSAT*, on the DIMACS 32-bit parity problem, which is one hard SAT problem in the DIMACS benchmarks [16]. The reason why we compared with only *EqSATz* is because so far it is the only solver which can solve all the ten par32* instances in reasonable time. Table 1 shows the run time of each solver on each instance in seconds. #var and #clause denotes the number of variables and the number of clauses in the input CNF formula, respectively. As shown in Table 1, Our solver, *XORSAT*, is able to solve all the ten par32* instances very fast. It can solve 7 of the 10 instances within one second. For this solver, the slowest is par32-3. Finding its solution took 3.235 seconds. For *EqSATz*, the slowest is par32-5. Solving it took 2844.6 seconds. This is $2844.6/2.922 \approx 974$ times slower than *XORSAT*, since the time required by *XORSAT* on this instance is 2.922 seconds.

## 6 Conclusions

The SAT solver introduced in this paper not only can solve all the ten par32* instances, but also is surprisingly fast. Its speed has been shown to be approximately 1000 times faster than the known fastest solver, *EqSATz*, on the DIMACS 32-bit parity problem, which is a challenge problem posed recently. We improved the performance of the solver in such a way: consider a SAT problem as a structured part and a random part, and then solve separately them. Our improvement is not engineering, but rather strategic. So the speedup is huge. We believe that this strategy is also useful beyond this domain.

Although this solver can solve all the 32-bit parity instances, we cannot conclude that it is complete. Thus, we have such an open problem: can it be proven theoretically to be complete? If it is proven to be incomplete, can it be improved to be complete? Also, how to integrate the strategy here with the Davis-Putnam method to speedup the SAT solver remains a future research work.